\documentclass[useAMS,usenatbib]{mn2e}

\usepackage{graphicx,amssymb,color}
\usepackage[normalem]{ulem}

\newcommand{\rev}{ }
\newcommand{\revv}{ }

\title[Debris environment of WD J0914+1914]
{The white dwarf planet WD J0914+1914~b: barricading potential rocky pollutants?}

\author[]{Dimitri Veras$^{1,2}$\thanks{E-mail: d.veras@warwick.ac.uk}\thanks{STFC Ernest Rutherford Fellow}
\\
$^{1}$Centre for Exoplanets and Habitability, University of Warwick, Coventry CV4 7AL, UK
\\
$^{2}$Department of Physics, University of Warwick, Coventry CV4 7AL, UK
}

\pubyear{2020}

\begin{document}
\label{firstpage}
\pagerange{\pageref{firstpage}--\pageref{lastpage}}
\maketitle

\begin{abstract}
An ice giant planet was recently reported orbiting white dwarf WD J0914+1914 at an approximate distance of 0.07 au. The striking non-detection of rocky pollutants in this white dwarf's photosphere contrasts with the observations of nearly every other known white dwarf planetary system. Here, I analyze the prospects for exterior extant rocky asteroids, boulders, cobbles and pebbles to radiatively drift inward past the planet due to the relatively high luminosity ($0.1 L_{\odot}$) of this particularly young (13~Myr) white dwarf. Pebbles and cobbles drift too slowly from Poynting-Robertson drag to bypass the planet, but boulders and asteroids are subject to the much stronger Yarkovsky effect. In this paper, I (i) place lower limits on the timescales for these objects to reach the planet's orbit, (ii) establish 3~m as the approximate limiting radius above which a boulder drifts too slowly to avoid colliding with the planet, and (iii) compute bounds on the fraction of boulders which succeed in traversing mean motion resonances and the planet's Hill sphere to eventually pollute the star. Overall, I find that the planet acts as a barrier against rather than a facilitator for radiatively-driven rocky pollution, suggesting that future rocky pollutants would most likely originate from distant scattering events.
\end{abstract}


\section{Introduction}

The most common signature of white dwarf planetary systems is contained within the atmospheres of the stars themselves: atoms heavier than helium \citep{vanmaanen1917,vanmaanen1919}, often referred to as {\it metals}. Nearly all of these metals should have sunk to the core soon after the birth of the white dwarf because of the star's high surface gravity (typically a factor of $10^5$ greater than the Earth's). Instead, the presence of these metals in over one quarter of Milky Way white dwarfs \citep{zucetal2003,zucetal2010,koeetal2014,couetal2019} implies ongoing accretion from planetary detritus.

Almost without exception \citep{xuetal2017}, the chemical nature of the debris is rocky, or volatile-poor \citep{ganetal2012,juryou2014,doyetal2019,swaetal2019}. Detailed analyses of the debris have linked it to potential formation locations of the progenitor minor planets \citep{haretal2018}, as well as the extent to which they were differentiated \citep{holetal2018,bonetal2020}. Reinforcing the rocky nature of this debris are transit-based \citep{vanetal2015,vanetal2019} and spectroscopic-based \citep{manetal2019} signatures of individual orbiting minor planets.

However, in none of these systems with rocky pollution has a major planet been detected so far. Strikingly then, \cite{ganetal2019} {\rev inferred the presence of} a likely ice giant planet orbiting the volatile-rich white dwarf WD J091405.30+191412.25 (henceforth WD J0914+1914). This {\rev finding} was based on the detection of abundant oxygen and sulfur (volatile elements) in combination with hydrogen throughout the system: both in the photosphere and the surrounding gaseous disc. The highly unusual (to the $10^{-4}$ level) abundant combination of these elements suggests that they arise from the deep layers of an ice giant planet -- layers which are being photoevaporated at a rate which is consistent with theoretical expectations \citep{schetal2019}.

No rocky material has been detected in this system -- yet -- although \cite{ganetal2019} derived upper limits for abundances of rocky elements in the white dwarf's photosphere. The planet's location is uncertain, but was modelled to reside at a distance of about 0.07 au from the star. Further, the cooling age of WD J0914+1914 (13 Myr) is 2-3 orders of magnitude younger than the oldest-known polluted white dwarfs, with cooling ages of about 8 Gyr \citep{holetal2017,holetal2018}. The combination of cooling age and inferred planet-star separation are challenging to reconcile theoretically without invoking the presence of at least one other giant planet in the system and the initiation of chaotic tidal circularization \citep{verful2019,verful2020}. In fact, \cite{verful2020} asserted that the planet, WD J0914+1914~b, is likely to be a ``puffy'' ice giant, and might be partially or even fully disrupted.

The youth of this system might imply that it is not yet dynamically settled \citep{veras2016}, particularly if WD J0914+1914~b was (plausibly) gravitationally scattered to its current location by another major planet \citep{debsig2002,veretal2013,voyetal2013,musetal2014,vergan2015,veretal2016,veretal2018}. Any moons that were stripped from these planets might still linger in the system \citep{payetal2016,payetal2017}, and unseen reservoirs of rocky debris from analogues of the Main Asteroid Belt or Kuiper Belt may be strewn about due to gravitational scattering with the major planets \citep{bonetal2011,debetal2012,frehan2014,antver2016,musetal2018,smaetal2018,antver2019}. Given this vast potential for rocky material to exist in the system, perhaps WD J0914+1914 is already polluted with rocky material, just below the detectable threshold.

Motivated by these varied scenarios, I wish to constrain the possibility of rocky pollution by modelling the limiting cases of inward radiative migration of rocky objects ranging in size from pebbles (mm-scale), and cobbles (cm- to dm-scale) to boulders (dm- to km-scale) and asteroids (km to Mm-scale) when coupled with the gravitational presence of WD J0914+1914~b\footnote{Rocky pollution could also arise from singular distant close encounters between unseen major and minor planets, where the latter is scattered directly towards the star while simply bypassing WD J0914+1914~b.}. 

Within this size range, Poynting-Robertson drag and the Yarkovsky effect (orbital recoil due to anisotropic re-emittance of absorbed radiation; see \citealt*{botetal2006} and \citealt*{voketal2015} for reviews) generate two types of radiative drift which influence different types of objects. \cite{veretal2015} demonstrated that radiative drag from the Yarkovsky effect is orders of magnitude stronger than Poynting-Robertson drag, but is unlikely to be activated for mm-scale pebbles nor cm-scale cobbles. Hence, I focus on boulders and asteroids, which could drift quickly enough to generate metal pollution. However, accurate modelling of the Yarkovsky effect requires detailed knowledge of the usually aspherical structure of the boulder or asteroid, and the subsequent recoil may be in any direction. Consequently, here I consider limiting cases only in order to provide definite and quantitative bounds.

In Section 2, I describe in more detail the Yarkovsky effect and the limiting case I adopt for my numerical simulations. In Section 3, I report the results of these simulations. I then discuss the results in Section 4 and conclude in Section 5.

\section{The maximum inward drag}

\subsection{Radiation effects with no planet}

An object orbiting WD J0914+1914 will absorb and reflect particular fractions of the incident white dwarf radiation; some of reflected radiation may occur after a delay, and in a different direction. The result is that the object's orbit, and potentially spin rate, will change. 

\cite{wyawhi1950} derived formulae which quantify the secular changes in semimajor axis and eccentricity when all of the incident radiation which is not absorbed is reflected immediately. The terminology for the evolution resulting from these formulae may be ambiguous \citep{buretal1979} but is commonly described as {\it Poynting-Robertson drag} \citep{poynting1904,robertson1937} and/or {\it radiation pressure}. For simplicity, I henceforth adopt the term Poynting-Robertson drag. Poynting-Robertson drag is always inward towards the radiation source.

If, instead, some of the incident radiation is redistributed within the object and subsequently is emitted anisotropically, then an additional drift is generated from what has become known as the {\it Yarkovsky effect} \citep{radzievskii1954,peterson1976}. This drift can be in any direction depending on the shape and mechanical properties of the object. Further, if the object is aspherical, then its spin rate will change in what has become known as the {\it YORP effect} \citep{rubincam2000}. A drastic consequence of spin rate change is rotational fission due to exceeding the break-up speed, a process which could have potentially generated abundant rocky debris when WD J0914+1914 was a giant branch star \citep{veretal2014a,versch2020}.

The Yarkovsky and YORP effects activate at different sizes. Depending on the object's thermal conductivity and thermal capacity, the Yarkovsky effect could become important in principle at arbitrarily low sizes; \cite{veretal2015} show that a reasonable lower bound is on the order of 0.1~m (the lower size end of what I {\rev henceforth} classify as boulders). They also illustrated that the Yarkovsky effect, at least during the giant branch phases of evolution, could significantly perturb asteroids as large as $10^3$ km.  For the critical size at which the YORP effect becomes significant, Solar system minor planet data provide constraints. For example, Fig. 1 of \cite{poletal2017} indicates that boulders with radii below about 100~m are preferentially not spun up to the point of destruction.

Concurrent modelling of Poynting-Robertson drag, the Yarkovsky effect and the YORP effect for arbitrarily-shaped objects on arbitrary orbits presents many challenges \citep[see e.g.][]{rozgre2012,rozgre2013}. \cite{veretal2015} simplified this situation by considering only the interplay between Poynting-Robertson drag and the Yarkovsky effect, and for spherical objects only (hence negating the YORP effect). Based on their results, I estimate that the drift induced by the Yarkovsky effect is up to a factor $c/(4\pi v_{\rm circ}) \sim 10^3$ greater than the drift induced by Poynting-Robertson drag, where $v_{\rm circ}$ is the circular speed of the object and $c$ is the speed of light.

The Yarkovsky drift may be in any direction, and would simultaneously change the semimajor axis, eccentricity and inclination of a boulder's orbit. Even if the boulder is considered to be a sphere, a full integration of the equations of motion would still be an onerous task, requiring calling matrices of spin and angular orbital momentum (see Eqs. 10-20 of \citealt*{veretal2015}) at each timestep of the integration. Instead, by considering a simplified treatment, one can bound the motion.

\subsection{Radiation effects with WD J0914+1914~b}

\cite{veretal2019} proceeded along this simplified route by fixing the values of these matrices while still retaining the relativistic direction correction term in its full generality.  They also simultaneously integrated the motion of a major planet along with Yarkovsky-induced asteroid drift. This setup is pertinent and relatable to WD J0914+1914, even though \cite{veretal2019} were integrating throughout the star's giant branch phase and with asteroids rather than the boulders I primarily consider here.

I hence adopt the same equations of motion as in \cite{veretal2019}, with the same assumptions, which I briefly reiterate here. The first assumption is that the boulder is spherical and has a constant density throughout its structure. This assumption both removes consideration of the YORP effect (which anyway would be negligible on the timescales and the sizes that I consider here) and allows me to characterize the boulders in terms of a well-defined radius. Also, I consider extreme cases of motion only, because I am placing limits on the evolution. Hence, I maximize the Yarkovsky effect by (i) assuming that all of the incoming radiation is absorbed by the boulder and then reflected after a delay (corresponding to diurnal and seasonal lag angles which together force the boulder inward; see equations 10-20 and 27-28 of \citealt*{veretal2015}), and (ii) equating the boulder's spin and orbital periods.

I establish the initial conditions to achieve maximum inward motion only. To do so, I adopt Model B from \cite{veretal2019} with perfectly coplanar and circular initial orbits. In this model B, the boulder's orbit would then remain coplanar throughout the evolution, allowing the semimajor axis to monotonically drift inward. As demonstrated in \cite{veretal2015}, the evolution of the semimajor axis, eccentricity and inclination of the boulder's orbit all depend on one another in nontrivial ways, but in limiting cases can be usefully characterized. I further reasonably assume that boulders are already external rather than internal to WD J0914+1914~b (which is located at 0.07 au), and so am not concerned about any directional drift other than directly inward: the purpose of the simulations is to establish limits on the motion. 

This inward drift will be differential with respect to the planet, which I assume maintains a fixed orbit. In this respect, the situation is roughly analogous to migration within a protoplanetary disc. \cite{veretal2019} considered in detail the conditions for capture into resonance and even into the Hill sphere of the planet (also see \citealt*{higida2016} and \citealt*{higida2017}). In the WD J0914+1914 system, my simulations will illustrate if and how boulders can evade capture into these resonances, and the planet's Hill sphere, to reach the white dwarf's photosphere.

\begin{figure*}
\centerline{
\includegraphics[width=16cm]{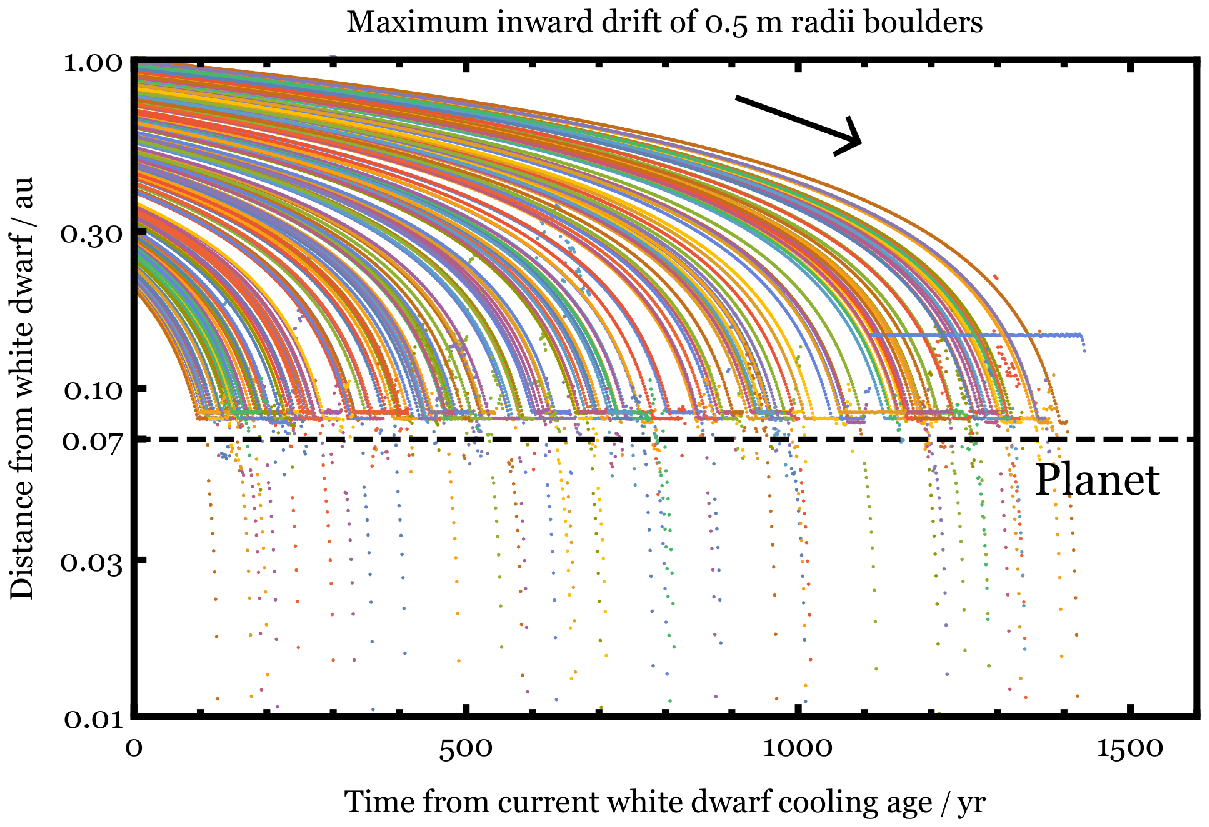}
}
\centerline{
\includegraphics[width=16cm]{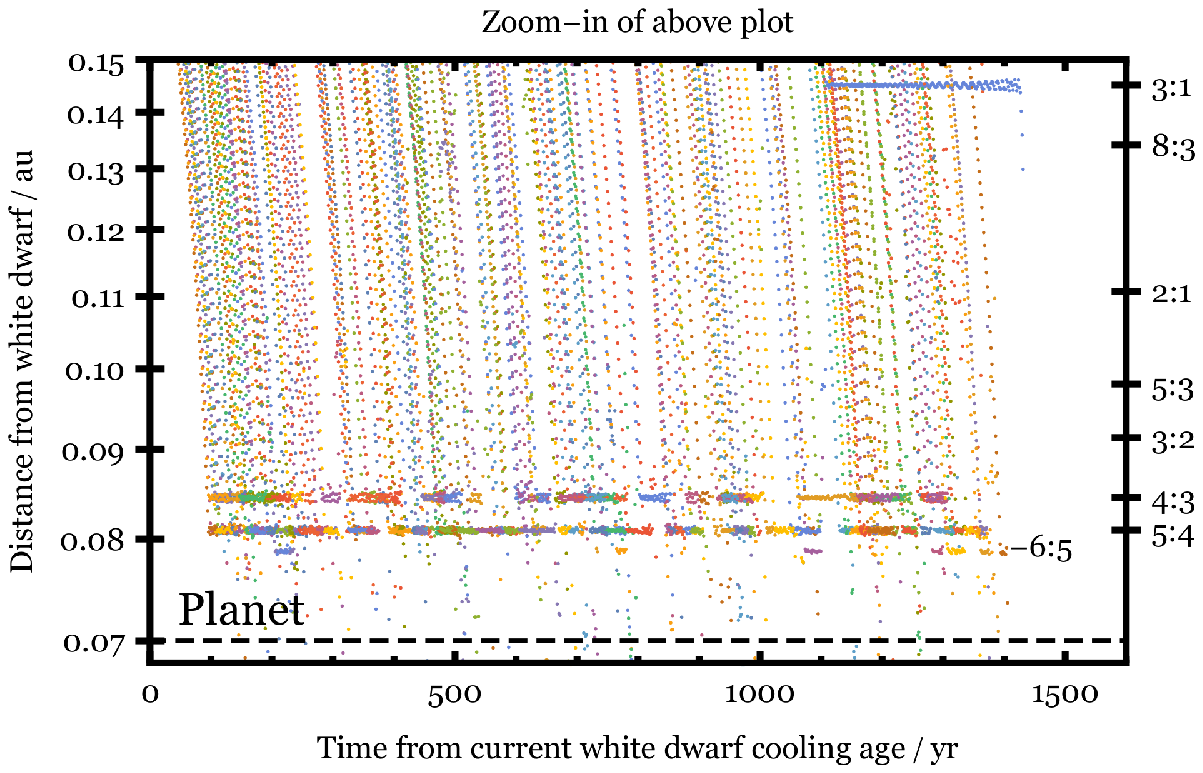}
}
\caption{Limiting inward drift of boulders driven by the Yarkovsky effect when it is activated and directed fully and constantly inward. The radii of these boulders is 0.5 m, and many are temporarily captured into mean motion resonances (bottom panel, right $y$-axis). Less than one quarter of all boulders avoid collision with the planet and instead pollute the white dwarf atmosphere. The output resolution is 1.5 yr, indicating how quickly some boulders are dragged into the white dwarf after bypassing the planet's orbit.
}
\label{cases}
\end{figure*}

\section{Simulations}

Now I describe my simulations. As suggested above, I used the same numerical code and equations of motion here that were presented in  \cite{veretal2019}. For the (variable timestep) RADAU integrator which I adopted, I set the accuracy parameter to be $10^{-11}$ for all integrations.

\subsection{Initial conditions}

I also adopted the parameters for WD J0914+1914 which were given in \cite{ganetal2019} (a mass of about $0.56M_{\odot}$, radius of about $0.015R_{\odot}$, and luminosity of about $0.1L_{\odot}$). Because my maximum integration timescales are $3 \times 10^4$ yr  (a few orders of magnitude smaller than the white dwarf cooling age of 13 Myr), I have kept the luminosity constant throughout the integrations. 

Further, investigations which consider planetary scattering in white dwarf systems commonly inflate the white dwarf radius to its Roche radius (which resides at about $1R_{\odot}$; see \citealt*{veretal2017} for a discussion of the subtleties of this estimate). Here, however, I used the true white dwarf radius for collision detection because (i) boulders may be small enough (depending on their material properties) to impact the white dwarf photosphere before fragmenting or sublimating \citep{broetal2017}, and (ii) once a boulder has reached $1R_{\odot}$, it cannot be ``saved'' and will eventually become a pollutant, even if delayed by first forming a ring or disc around the white dwarf \citep{debetal2012,veretal2014b,malper2020a,malper2020b}. 

Given the close proximity of some of these boulders to the white dwarf {\rev and the possibility of their orbital eccentricities becoming nonzero when interacting with the planet}, I also included the effects of general relativity in the integrations. {\rev General relativity generates secular changes in the evolution of the argument of pericentre, and non-secular changes on the scale of up to km in pericentre distance on a per-orbit basis \citep{veras2014}. However, my results will demonstrate that general relativity likely had a negligible effect on the evolution.}

The physical properties of WD J0914+1914~b are unconstrained observationally, except for the chemical match to ice giants. From the theoretical perspective, \cite{verful2020} provided coupled sets of constraints on the planet's physical properties and type of the planet's orbit. For simplicity and definitiveness, and because of computational constraints, I adopted a Neptune-mass for WD J0914+1914~b in all simulations. I also assumed that it resides on a circular orbit of 0.07 au. As previously mentioned, in order to generate limiting inward motion from the Yarkovsky effect, I set the planet's orbit to be coplanar with those of the boulders. 

Although the boulders were modelled to not gravitationally perturb one another nor the planet, they are not massless: they required a finite radius in order to experience radiative drag. In order to determine what radii would be relevant (the answer being up to 10 m, which is why I predominantly use the term {\it boulder}), I first performed a preliminary sparse set of simulations. These initial simulations also helped established the maximum initial separation of the boulders for which the numerical integrations took no longer than a couple weeks of CPU time (up to 1 au for 10 m boulders, and up to 5 au for 0.1 m boulders), and hence the maximum necessary duration of the simulations ($3 \times 10^4$ yr) such that all boulders reach the planet's orbit. In all cases, I assumed the boulder density to be $2$ g/cm$^3$; the results are relatively independent of this choice (e.g. increasing the density to $5$ g/cm$^3$ would decrease the effective radius by only 27 per cent, for a fixed mass).

Like the planet, the boulders have orbits which I initialized to be circular. Crucially, the relative orientation of an individual boulder and the planet (given by their mean or true anomalies) determines if the boulder would eventually be captured into a mean motion resonance with the planet. Predicting exactly which of these initial values leads to mean motion resonance capture {\it a priori} would require a detailed analysis, and may not even be possible, especially given that the eccentricities of the boulders' orbits immediately become non-zero (see \citealt*{veretal2019} for a discussion). Hence, I randomly sampled the initial mean anomalies of the boulders' orbits from a uniform distribution.

I sampled eight different boulder radii (0.1, 0.5, 1.0, 2.0, 3.0, 4.0, 5.0, 10 m) and six different initial boulder-star separation ranges (0.2-0.4, 0.4-0.6, 0.6-0.8, 0.8-1.0, 1.0-2.0 and 2.0-5.0 au). The number of boulders I simulated varied nonuniformly across this parameter space; for a given radius and separation range, I sampled between about 150 and 600 different boulders. I set the data output frequency to be $0.1$ per cent of the duration of the simulation, equating to a output frequency range of six months to 30 years depending on the individual simulation.

\begin{figure*}
\centerline{
\includegraphics[width=16cm]{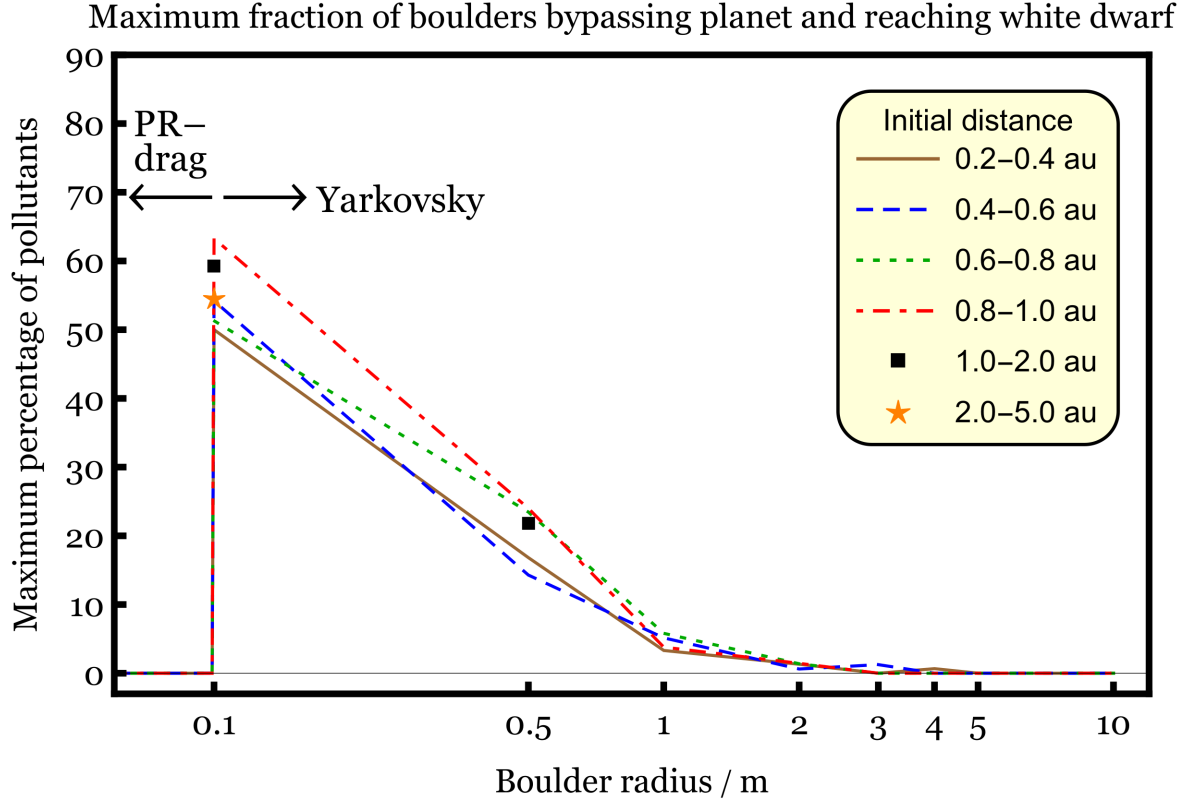}
}
\caption{
Demonstration that radiatively-driven rocky pollutants can arise only within the boulder radii range of about 0.1~m to 3~m. This lower size bound depends on the material properties of the boulders and could be higher or lower, while the upper bound is limited by the resolution of the simulations (a total of 600 boulders were simulated for each radius above 2~m). In the regime where Poynting-Robertson drag dominates, the cobbles and pebbles move too slowly to bypass the planet and pollute the white dwarf. Also, all asteroids (with radii of at least 1 km) are too large to pollute the white dwarf.
}
\label{punch}
\end{figure*}

\subsection{Results}

\begin{figure*}
\centerline{
\includegraphics[width=16cm]{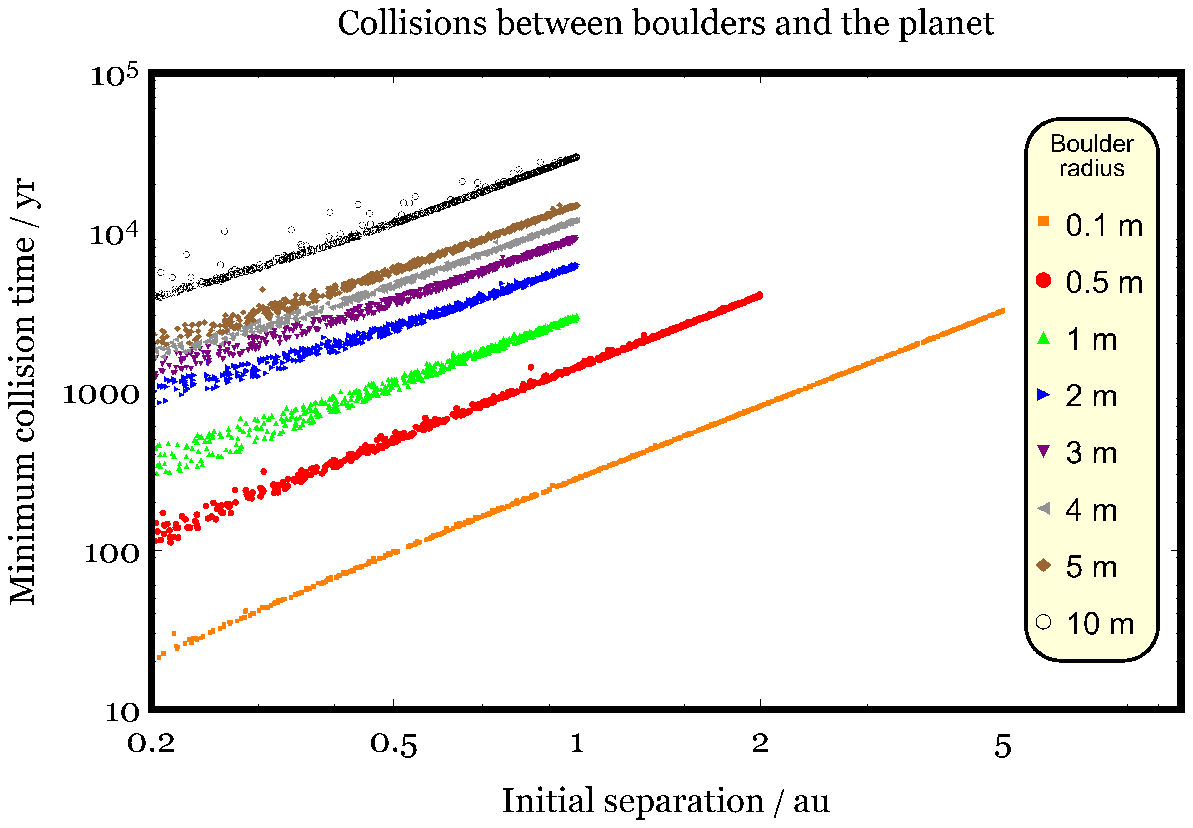}
}
\caption{
Time for a boulder with a given radius and initial separation to collide with the planet; only collision events with the planet are shown. For any given boulder radius, although some scatter is apparent in the collision times due to temporary capture in mean motion resonance, the collision times follow robust trends which have predictive power {\rev (see equation \ref{fitline})}.
}
\label{collision}
\end{figure*}

My results are summarized in Figs.~\ref{cases}-\ref{collision}. Figure~\ref{cases} displays one representative example of a numerical simulation, whereas Figs.~\ref{punch}-\ref{collision} instead provide ensembles of the results; Fig.~\ref{punch} expresses the main result.

\subsubsection{Figure~\ref{cases}: sample evolution}

Figure ~\ref{cases} illustrates the evolution of 0.5 m radii boulders drifting towards WD~J0914+1914~b. Some boulders bypass the planet orbit at 0.07 au (proceeding to pollute the white dwarf), but most do not, and instead collide with the planet. On their way towards the planet, many of the boulders are temporarily trapped in mean motion resonances. The bottom panel of the figure best illustrates this behaviour. 

That panel reveals that in every case, capture is only temporary. The reason is because the Yarkovsky effect eventually wrenches the boulder out of the resonance. The panel also reveals that the most populated mean motion resonances correspond to the $4$:$3$ and $5$:$4$ commensurabilities, even though the $2$:$1$ and $3$:$2$ resonances are comparable in strength. This result directly follows from the inward speed of the boulders, and is easily adjusted by changing the radius of the boulder or luminosity of the central star. For example, temporary $2$:$1$ resonance capture is predominant in the 5 m and 10 m radius boulder simulations (not shown). 

I also note the temporary capture of one boulder into the weaker, second-order $3$:$1$ mean motion resonance. This standout feature on the bottom panel of the figure {\rev (in light blue)} exemplifies how finely-tuned orbital parameters of the boulder and planet could lead to capture in a wide variety of resonances. The increasing amplitude with time of this feature also nicely illustrates gradual escape from resonance. Further, resonances could have an effect on the motion even if they don't generate capture; I displayed the location of the (very weak) $8$:$3$ resonance because it creates a noticeable disturbance, or inhomogeneity, on a boulder's evolution at about 500 yr into the simulation. I finally note that if the Yarkovsky effect ``turns off'' while a boulder is in resonance (perhaps due to a physical change in the boulder), then it may remain in that resonance.

\subsubsection{Figure~\ref{punch}: pollution statistics}

The primary goal of this paper is to determine if radiative drag from the young and luminous WD~J0914+1914 could drive rocky pollution from extant rocky objects exterior to the planet. Figure ~\ref{punch} indicates that the contribution can arise only in a narrow range of boulder radii whose upper limit is about 3 m and whose lower limit is the (arguable) activation size for the Yarkovsky effect. Further, even within this range, the Yarkovsky effect must be oriented favourably, as I am only presenting the limiting, maximally polluting case.

The upper limit of 3 m may also represent a function of the resolution of my simulations. In this respect, I can claim that no more than about 0.1 per cent of boulders of larger radii could pollute the white dwarf. 

The lower radius limit of 0.1 m is highly dependent on the material properties and spin that are assumed for the boulder. Below this limit, cobbles and pebbles, which would be subject to Poynting-Robertson drag only, would move too slowly to pollute the white dwarf: their fate would be similar to asteroids or boulders larger than 10 m moving at comparable speeds, and hence do not need to be simulated\footnote{Because white dwarfs have no winds, cobbles and pebbles cannot be blown out of the system, as they might have been when WD~J0914+1914 was a giant branch star \citep{zotver2020}.}.

Figure ~\ref{punch} also illustrates that the pollution fraction does not correlate with initial distance. The reason is because all boulders will approach the planet's orbit at approximately the same speed, regardless of starting location. Further, the curves in the figure may be directly scaled to different luminosities corresponding to different white dwarf cooling ages: when WD~J0914+1914 was younger, pollution from larger boulders may have been more prevalent.

\subsubsection{Figure~\ref{collision}: collision times}

Another outcome of this study are bounds on the inward migration timescale of the boulders. Plotted on Fig.~\ref{collision} are the limiting collision timescales with the planet as a function of initial separation and boulder radius. Although individual simulations are plotted, the trends are robust and distinctive enough to appear as lines for increasing initial separation. These results demonstrate predictive power, and may be scaled to different cooling ages or boulder radii which I did not simulate.

{\rev In fact, by assuming that the averaged semimajor axis evolution is a proxy for inward drift ({\revv thereby} neglecting eccentricity), I can obtain an analytical scaling for the symbols in Fig.~\ref{collision}. Equation (103) of \cite{veretal2015} suggests that

\begin{equation}
t_{\rm col} \propto L_{\star}^{-1} M_{\star}^{\frac{1}{2}} \rho R a_{\rm i}^{\frac{3}{2}}
,
\end{equation}

\noindent{}where $t_{\rm col}$ is the minimum time for a boulder to collide with WD J0914+1914~b, $L_{\star}$ and $M_{\star}$ are the luminosity and mass of the white dwarf, and $\rho$, $R$ and $a_{\rm i}$ are the density, radius and initial semimajor axis of the boulder. When this relation is fitted to Fig.~\ref{collision}, I obtain

\[
t_{\rm col} \approx 
1 \times 10^4 \ {\rm yr} 
\left( \frac{L_{\star}}{0.1 L_{\odot}} \right)^{-1}
\left( \frac{M_{\star}}{0.56 M_{\odot}} \right)^{\frac{1}{2}}
\]

\begin{equation}
\ \ \ \ \times \left( \frac{\rho}{2 \ {\rm g}/{\rm cm}^3} \right)
\left( \frac{R}{4 \ {\rm m}} \right)
\left( \frac{a_{\rm i}}{1 \ {\rm au}} \right)^{\frac{3}{2}}
.
\label{fitline}
\end{equation}

}

{\rev Not all symbols follow the power-law suggested by equation (\ref{fitline}). These outliers} are due to relatively long captures in mean motion resonances with the planet. The scatter in collision times is greatest for the 10~m simulations, perhaps resulting from easy and lengthy capture into the strong $2$:$1$ resonance due to the slow speed of these boulders.

\section{Discussion}

I have shown that WD~J0914+1914 has the potential to be polluted with radiatively-driven rocky material which span only about one order of magnitude in radius. This restriction on the size distribution would likely stand out in the results of \cite{wyaetal2014}, who linked observed accretion rates to different size distributions of pollutants and whether that accretion was stochastic or continuous. For the known population of metal-polluted white dwarfs at the time,  \cite{wyaetal2014} effectively ruled-out accretion from a mono-mass distribution, which would contrast with the available radiatively-driven pollutants in WD~J0914+1914. The pollution would, however, likely be consistent with the finding that accretion from a single object is unlikely \citep{turwya2019}.

In the future, WD~J0914+1914 may still be polluted non-radiatively through distant scattering events from unseen minor and major planets which have survived the giant branch phases of stellar evolution. Bodies perturbed towards the white dwarf would be gravitationally focussed by WD~J0914+1914~b, which highlights another difference from other white dwarfs without tight-orbit major planets where minor planets would have an unimpeded path to the white dwarf Roche radius. 

A particularly pertinent unknown quantity is the mass of WD~J0914+1914~b. \cite{verful2020} suggested that the planet may be partially or fully evaporated. A less massive planet than what I modelled here would facilitate radiatively-driven rocky pollution. Reducing the planet's mass would also alter the resonant structure of the system, and ultimately lead to fewer collisions with the remnant of the planet (and more pollution onto the white dwarf).

This evaporation likely fed and continues to feed the presence of a gas disc. \cite{ganetal2019} found that WD~J0914+1914 harbours a circumstellar gas disc with inner and outer radii of approximately 0.005 au and 0.05 au. A striking feature of this disc is that it represents the first known white dwarf debris disc that does not contain dust {\rev \citep{zucbec1987,ganetal2006,farihi2016,denetal2018,manetal2020}}, at least at a currently detectable level. 

The gas would create drag on radiatively-driven boulders which bypass the planet, thereby delaying -- but not preventing -- the accretion of this material onto the white dwarf photosphere. Depending on the physical properties of the boulder, it might even be sublimated before reaching the white dwarf photosphere \citep{broetal2017}. Future modelling the disc structure subject to the gravity of WD~J0914+1914~b would be desirable. Further, coupling this structure with evolution due to incoming boulders would be well-suited for numerical cascade codes such as the one presented in \cite{kenbro2017}.

\section{Conclusion}

WD~J0914+1914~b is the first major planet reported orbiting a single white dwarf on a close orbit (0.07 au). The lack of rocky pollutants in this white dwarf's photosphere is a striking and almost unique feature of the known population of white dwarf planetary systems. Motivated by (i) this anomaly, (ii) the fact that WD~J0914+1914 is such a young (13 Myr) and luminous ($0.1L_{\odot}$) white dwarf, and (iii) the likely possibility that gravitational scattering and dynamical re-arrangement already occurred in this system, in this paper I explored the possibility of radiatively-driven rocky pollution from extant debris residing exterior to the planet. 

I found that while pebbles, cobbles and asteroids cannot pollute the white dwarf {\rev through radiatively-driven migration}, the Yarkovsky effect could speed up boulders with a maximum radius of approximately 3 m to sufficiently high to levels to bypass mean motion resonances with the planet (Fig.~\ref{cases}), as well as the planet's Hill sphere itself, to pollute the white dwarf. However, I modelled only the bounding, idealized case of continuous inward migration, concluding that the efficacy of this process (Fig.~\ref{punch}) is limited. The migration timescale for boulders to reach the planet is relatively unaffected by the latter's presence, and can be scaled to different boulder radii and white dwarf cooling ages (Fig.~\ref{collision} and {\rev equation \ref{fitline}}). As WD~J0914+1914 cools, the possibility of radiatively-driven rocky pollution will gradually disappear because incoming boulders will instead all collide with the planet.

\section*{Acknowledgements}

{\rev I thank the referee for helpful comments which have improved the manuscript.} I also gratefully acknowledge the support of the STFC via an Ernest Rutherford Fellowship (grant ST/P003850/1).

\label{lastpage}

\begin{thebibliography}{99}

\bibitem[Antoniadou \& Veras(2016)]{antver2016} Antoniadou, K.~I., \& Veras, D.\ 2016, MNRAS, 463, 4108 

\bibitem[Antoniadou \& Veras(2019)]{antver2019} Antoniadou, K.~I., \& Veras, D.\ 2019, A\&A, 629, A126

\bibitem[Bonsor et al.(2011)]{bonetal2011} Bonsor, A., Mustill, A.~J., \& Wyatt, M.~C.\ 2011, MNRAS, 414, 930 

\bibitem[Bonsor et al.(2020)]{bonetal2020} Bonsor, A., Carter, P.~J., Hollands, M., et al.\ 2020, MNRAS, 3235

\bibitem[Bottke et al.(2006)]{botetal2006} Bottke, W.~F., Vokrouhlick{\'y}, D., Rubincam, D.~P., et al.\ 2006, Annual Review of Earth and Planetary Sciences, 34, 157

\bibitem[Brown et al.(2017)]{broetal2017} Brown, J.~C., Veras, D., \& G{\"a}nsicke, B.~T.\ 2017, MNRAS, 468, 1575 

\bibitem[Burns et al.(1979)]{buretal1979} Burns, J.~A., Lamy, P.~L., \& Soter, S.\ 1979, Icarus, 40, 1

\bibitem[Coutu et al.(2019)]{couetal2019} Coutu, S., Dufour, P., Bergeron, P., et al.\ 2019, ApJ, 885, 74

\bibitem[Debes \& Sigurdsson(2002)]{debsig2002} Debes, J.~H., \& Sigurdsson, S.\ 2002, ApJ, 572, 556 

\bibitem[Debes et al.(2012)]{debetal2012} Debes, J.~H., Walsh, K.~J., \& Stark, C.\ 2012, ApJ, 747, 148 

\bibitem[Dennihy et al.(2018)]{denetal2018} Dennihy, E., Clemens, J.~C., Dunlap, B.~H., Fanale, S.~M., Fuchs, J.~T., Hermes, J.~J.\ 2018, ApJ, 854, 40 

\bibitem[Doyle et al.(2019)]{doyetal2019} Doyle, A.~E., Young, E.~D., Klein, B., et al.\ 2019, Science, 366, 356

\bibitem[Farihi(2016)]{farihi2016} Farihi, J.\ 2016, New Astronomy Reviews, 71, 9 

\bibitem[Frewen \& Hansen(2014)]{frehan2014} Frewen, S.~F.~N., \& Hansen, B.~M.~S.\ 2014, MNRAS, 439, 2442 

\bibitem[G{\"a}nsicke et al.(2006)]{ganetal2006} G{\"a}nsicke, B.~T., Marsh, T.~R., Southworth, J., \& Rebassa-Mansergas, A.\ 2006, Science, 314, 1908 

\bibitem[G{\"a}nsicke et al.(2012)]{ganetal2012} G{\"a}nsicke, B.~T., Koester, D., Farihi, J., et al.\ 2012, MNRAS, 424, 333 

\bibitem[G{\"a}nsicke et al.(2019)]{ganetal2019} G{\"a}nsicke, B.~T., Schreiber, M.~R., Toloza, O., et al.\ 2019, Nature, 576, 61

\bibitem[Harrison et al.(2018)]{haretal2018} Harrison, J.~H.~D., Bonsor, A., \& Madhusudhan, N.\ 2018, MNRAS, 479, 3814.

\bibitem[Higuchi \& Ida(2016)]{higida2016} Higuchi, A., \& Ida, S.\ 2016, AJ, 151, 16

\bibitem[Higuchi \& Ida(2017)]{higida2017} Higuchi, A., \& Ida, S.\ 2017, AJ, 153, 155

\bibitem[Hollands et al.(2017)]{holetal2017} Hollands, M.~A., Koester, D., Alekseev, V., Herbert, E.~L., \& G{\"a}nsicke, B.~T.\ 2017, MNRAS, 467, 4970 

\bibitem[Hollands et al.(2018)]{holetal2018} Hollands, M.~A., G{\"a}nsicke, B.~T., \& Koester, D.\ 2018, MNRAS, 477, 93.

\bibitem[Jura \& Young(2014)]{juryou2014} Jura, M., \& Young, E.~D.\ 2014, Annual Review of Earth and Planetary Sciences, 42, 45 

\bibitem[Kenyon \& Bromley(2017)]{kenbro2017} Kenyon, S.~J., \& Bromley, B.~C.\ 2017, ApJ, 850, 50

\bibitem[Koester et al.(2014)]{koeetal2014} Koester, D., G{\"a}nsicke, B.~T., \& Farihi, J.\ 2014, A\&A, 566, A34 

\bibitem[Malamud \& Perets(2020a)]{malper2020a} Malamud, U., \& Perets, H.\ 2020a, arXiv:1911.12068

\bibitem[Malamud \& Perets(2020b)]{malper2020b} Malamud, U., \& Perets, H.\ 2020b, arXiv:1911.12184

\bibitem[Manser et al.(2019)]{manetal2019} Manser, C.~J., G{\"a}nsicke, B.~T., Eggl, S., et al.\ 2019, Science, 364, 66

\bibitem[Manser et al.(2020)]{manetal2020} Manser, C.~J., G{\"a}nsicke, B.~T., Gentile Fusillo, N.~P., et al.\ 2020, MNRAS In Press, arXiv:2002.01936

\bibitem[Mustill et al.(2014)]{musetal2014} Mustill, A.~J., Veras, D., \& Villaver, E.\ 2014, MNRAS, 437, 1404 

\bibitem[Mustill et al.(2018)]{musetal2018} Mustill, A.~J., Villaver, E., Veras, D.,  G{\"a}nsicke, B.~T., Bonsor, A. \ 2018, MNRAS, 476, 3939.

\bibitem[Payne et al.(2016)]{payetal2016} Payne, M.~J., Veras, D., Holman, M.~J., G\"{a}nsicke, B.~T.\ 2016, MNRAS, 457, 217 

\bibitem[Payne et al.(2017)]{payetal2017} Payne, M.~J., Veras, D., G{\"a}nsicke, B.~T., \& Holman, M.~J.\ 2017, MNRAS, 464, 2557 

\bibitem[Peterson(1976)]{peterson1976} Peterson, C.\ 1976, Icarus, 29, 91

\bibitem[Polishook et al.(2017)]{poletal2017} Polishook, D., Moskovitz, N., Thirouin, A., et al.\ 2017, Icarus, 297, 126

\bibitem[Poynting(1904)]{poynting1904} Poynting, J.~H.\ 1904, Philosophical Transactions of the Royal Society of London Series A, 202, 525

\bibitem[Radzievskii(1954)]{radzievskii1954} Radzievskii V. V., 1954, Dokl. Akad Nauk SSSR, 97, 49

\bibitem[Robertson(1937)]{robertson1937} Robertson, H.~P.\ 1937, MNRAS, 97, 423

\bibitem[Rozitis \& Green(2012)]{rozgre2012} Rozitis, B., \& Green, S.~F.\ 2012, MNRAS, 423, 367

\bibitem[Rozitis \& Green(2013)]{rozgre2013} Rozitis, B., \& Green, S.~F.\ 2013, MNRAS, 433, 603

\bibitem[Rubincam(2000)]{rubincam2000} Rubincam, D.~P.\ 2000, Icarus, 148, 2

\bibitem[Schreiber et al.(2019)]{schetal2019} Schreiber, M.~R., G{\"a}nsicke, B.~T., Toloza, O., et al.\ 2019, ApJL, 887, L4

\bibitem[Smallwood et al.(2018)]{smaetal2018} Smallwood, J.~L., Martin, R.~G., Livio, M., \& Lubow, S.~H.\ 2018, MNRAS, 480, 57

\bibitem[Swan et al.(2019)]{swaetal2019} Swan, A., Farihi, J., Koester, D., et al.\ 2019, MNRAS, 490, 202

\bibitem[Turner \& Wyatt(2019)]{turwya2019} Turner, S.~G.~D., \& Wyatt, M.~C.\ 2019, MNRAS, 2788

\bibitem[van Maanen(1917)]{vanmaanen1917} van Maanen, A.\ 1917, PASP, 29, 258 

\bibitem[van Maanen(1919)]{vanmaanen1919} van Maanen, A.\ 1919, AJ, 32, 86 

\bibitem[Vanderbosch et al.(2019)]{vanetal2019} Vanderbosch, Z., Hermes, J.~J., Dennihy, E., et al.\ 2019, Submitted to ApJL, arXiv:1908.09839

\bibitem[Vanderburg et al.(2015)]{vanetal2015} Vanderburg, A., Johnson, J.~A., Rappaport, S., et al.\ 2015, Nature, 526, 546 

\bibitem[Veras et al.(2013)]{veretal2013} Veras, D., Mustill, A.~J., Bonsor, A., \& Wyatt, M.~C.\ 2013, MNRAS, 431, 1686 

\bibitem[Veras(2014)]{veras2014} Veras, D.\ 2014, MNRAS, 442, L71

\bibitem[Veras et al.(2014a)]{veretal2014a} Veras, D., Jacobson, S.~A., G\"{a}nsicke, B.~T.\ 2014a, MNRAS, 445, 2794 

\bibitem[Veras et al.(2014b)]{veretal2014b} Veras, D., Leinhardt, Z.~M., Bonsor, A., G\"{a}nsicke, B.~T.\ 2014b, MNRAS, 445, 2244

\bibitem[Veras \& G\"{a}nsicke(2015)]{vergan2015} Veras, D., G\"{a}nsicke, B.~T.\ 2015, MNRAS, 447, 1049 

\bibitem[Veras et al.(2015)]{veretal2015} Veras, D., Eggl, S., G{\"a}nsicke, B.~T.\ 2015, MNRAS, 451, 2814 

\bibitem[Veras(2016)]{veras2016} Veras, D.\ 2016, Royal Society Open Science, 3, 150571 


\bibitem[Veras et al.(2016)]{veretal2016} Veras, D., Mustill, A.~J., G{\"a}nsicke, B.~T., et al.\ 2016, MNRAS, 458, 3942 

\bibitem[Veras et al.(2017)]{veretal2017} Veras, D., Carter, P.~J., Leinhardt, Z.~M., et al.\ 2017, MNRAS, 465, 1008

\bibitem[Veras et al.(2018)]{veretal2018} Veras D., Georgakarakos N., G{\"a}nsicke B.~T., Dobbs-Dixon I., 2018a, MNRAS, 481, 2180

\bibitem[Veras et al.(2019)]{veretal2019} Veras, D., Higuchi, A., \& Ida, S.\ 2019, MNRAS, 485, 708

\bibitem[Veras \& Fuller(2019)]{verful2019} Veras, D., \& Fuller, J.\ 2019, MNRAS, 489, 2941

\bibitem[Veras \& Fuller(2020)]{verful2020} Veras, D., \& Fuller, J.\ 2020, MNRAS, 492, 6059

\bibitem[Veras \& Scheeres(2020)]{versch2020} Veras, D., \& Scheeres, D.~J.\ 2020, MNRAS, 492, 2437

\bibitem[Vokrouhlick{\'y} et al.(2015)]{voketal2015} Vokrouhlick{\'y}, D., Bottke, W.~F., Chesley, S.~R., et al.\ 2015, In Asteroids IV (Eds. P. Michel, F.~E. DeMeo and W.~F. Bottke), 509

\bibitem[Voyatzis et al.(2013)]{voyetal2013} Voyatzis, G., Hadjidemetriou, J.~D., Veras, D., \& Varvoglis, H.\ 2013, MNRAS, 430, 3383 

\bibitem[Wyatt et al.(2014)]{wyaetal2014} Wyatt, M.~C., Farihi, J., Pringle, J.~E., \& Bonsor, A.\ 2014, MNRAS, 439, 3371 

\bibitem[Wyatt \& Whipple(1950)]{wyawhi1950} Wyatt, S.~P., \& Whipple, F.~L.\ 1950, ApJ, 111, 134

\bibitem[Xu et al.(2017)]{xuetal2017} Xu, S., Zuckerman, B., Dufour, P., et al.\ 2017, ApJL, 836, L7 

\bibitem[Zotos \& Veras(2020)]{zotver2020} Zotos, E., \& Veras, D.\ 2020, Submitted to A\&A

\bibitem[Zuckerman \& Becklin(1987)]{zucbec1987} Zuckerman, B., \& Becklin, E.~E.\ 1987, Nature, 330, 138

\bibitem[Zuckerman et al.(2003)]{zucetal2003} Zuckerman, B., Koester, D., Reid, I.~N., H\"{u}nsch, M.\ 2003, ApJ, 596, 477 

\bibitem[Zuckerman et al.(2010)]{zucetal2010} Zuckerman, B., Melis, C., Klein, B., Koester, D., \& Jura, M.\ 2010, ApJ, 722, 725 


\end{thebibliography}
\end{document}